\documentclass[12pt,aip,apl]{revtex4-1}
\usepackage{graphicx} % For including images
\usepackage{xcolor}
\usepackage{color,soul}%to highlight text
\usepackage{caption}
\usepackage{siunitx}
\usepackage{amssymb}   % For additional math symbols (e.g. \mathbb, \leqslant)
\usepackage{amsfonts}  % For math fonts (e.g. \mathbb, \mathfrak)
\usepackage{mathtools} % Improvements to amsmath
\usepackage[utf8]{inputenc}
\usepackage[colorlinks=true,urlcolor=blue,linkcolor=black,citecolor=black]{hyperref} % Hyperlinks with color
\usepackage[utf8]{inputenc}
\newcommand{\NdIII}{\ensuremath{\text{Nd}^{3+}}}
\newcommand{\YbIII}{\ensuremath{\text{Yb}^{3+}}}
\usepackage[thinc]{esdiff}
\usepackage{nicefrac}
\usepackage{lineno}

\begin{document}

%\preprint{}
\title{Solar-pumped Radiation-balanced Laser}


\author{Michael Küblböck}
\affiliation{Max Planck Institute for the Science of Light, Staudtstr 2, Erlangen, 91058, Germany}%
\affiliation{Friedrich-Alexander-Universität Erlangen-Nürnberg, Staudstr 7, Erlangen, 91058, Germany}%
\affiliation{Graduate School in Advanced Optical Technologies (SAOT), Konrad-Zuse-Straße 3, Erlangen, 91052, Germany}%

\author{Mohammad Sahil}
\affiliation{Max Planck Institute for the Science of Light, Staudtstr 2, Erlangen, 91058, Germany}%
\affiliation{Friedrich-Alexander-Universität Erlangen-Nürnberg, Staudstr 7, Erlangen, 91058, Germany}%

%\affiliation{SAOT, Staudstrasse 7, Erlangen, 91058, Germany}%
\author{Jasvinder Brar}
\affiliation{Max Planck Institute for the Science of Light, Staudtstr 2, Erlangen, 91058, Germany}%
\affiliation{Friedrich-Alexander-Universität Erlangen-Nürnberg, Staudstr 7, Erlangen, 91058, Germany}%

\author{Hanieh Fattahi}%
 \email{hanieh.fattahi@mpl.mpg.de}
\affiliation{Max Planck Institute for the Science of Light, Staudtstr 2, Erlangen, 91058, Germany}%
\affiliation{Friedrich-Alexander-Universität Erlangen-Nürnberg, Staudstr 7, Erlangen, 91058, Germany}%

\begin{abstract}
Solar-pumped lasers, predominantly based on neodymium gain media, offer a promising route to renewable laser--energy conversion and space-based photonics; however, their performance has been constrained by thermal loading and limited power scalability. Here, we propose and numerically investigate a solar-pumped ytterbium thin-disk gain medium combined with a dome concentrator, which enables multipass solar pumping and enhanced absorption. The design yields comparably low lasing thresholds for neodymium- and ytterbium-doped media, while ytterbium provides superior power scalability, enabling up to threefold higher output power. We further identify ytterbium-doped medium combined with a spherical concentrator as a viable solar-pumped, radiation-balanced configuration, achieving self-cooled lasing at solar pump intensities of $28.5~\mathrm{kW\,cm^{-2}}$ within the $1020$--$1033~\mathrm{nm}$ window of the solar spectrum. We further demonstrate that dual-wavelength pumping overcomes the limitations imposed by low solar intensity and concentration constraints, enabling radiation-balanced lasing at orders-of-magnitude lower solar pump intensities. The proposed spherical-concentrator-based design enhances pump absorption while allowing efficient escape of anti-Stokes fluorescence. These results establish multi-pass, solar-pumped ytterbium lasers as a compact, scalable, and sustainable platform for high-performance solar-pumped lasers.
\end{abstract}

\maketitle

\section*{Introduction}

Solar-pumped lasers have gained significant attention over the past decade as a sustainable technology with potential applications in renewable energy, green laser systems, communications, and space exploration, particularly for future lunar infrastructure \cite{kublbock2024solar, national2007space, algora2022beaming, Hemmati:07, edenhofer2011renewable, mackay2016sustainable, brandon2021role, osman2022hydrogen}. Most demonstrated systems employ solid-state rod lasers that harness the sun’s abundant energy and convert it directly into tens of watts of coherent optical power. However, their overall efficiency remains constrained by the substantial thermal load imposed on the gain medium.

The first continuous-wave solid-state solar laser, demonstrated by Young, utilized a neodymium-doped yttrium aluminum garnet (Nd:YAG) crystal as the gain medium \cite{young1966sun}. \NdIII, a well-known four-level laser ion, offers a low lasing threshold \cite{garcia2022nd, do2011, liang2012}. Its emission cross-section bandwidth of approximately \SI{0.6}{\nm} enables the generation of nanosecond-scale pulses. In Nd:YAG systems, however, the \NdIII\ doping concentration is typically limited to approximately \SI{1.1}{\%} due to fluorescence quenching, which reduces the fluorescence lifetime at higher dopant levels. This behavior is intrinsic to \NdIII\ and arises from cross-relaxation within the $^4F_{3/2}$ manifold as well as near-field electric-dipole interactions between adjacent \NdIII\ ions under continuous pumping \cite{danielmeyer1973fluorescence, Merkle:06}. At lower doping levels, a longer optical path within the gain medium is required to ensure sufficient pump absorption. As pump power increases, thermal management becomes a critical design constraint. Excess heat generated by quantum defects and non-radiative decay processes produces temperature gradients within the crystal, leading to transient thermal lensing and beam distortion. These effects exacerbate non-radiative losses and induce depolarization \cite{foster1970thermal, Stone:92, eichler1993thermal, Koechner:70, Guan2018}.

In recent years, co-doping strategies have been explored to improve the performance of neodymium-based solar-pumped lasers \cite{berwal2022review, kublbock2024solar, Payziyev2019, liang2012, Vistas.2022, Vistas.2021, Vistas.2020, Liang.2018, Liang.2013, Liang.2013b}. Co-doping Nd$^{3+}$ with sensitizer ions such as Cr$^{3+}$ or Ce$^{3+}$ enhances spectral overlap with the solar spectrum, enabling more efficient absorption and energy transfer, and thus improving pump efficiency and laser output. However, high sensitizer concentrations can increase thermal load, scattering losses, and lattice distortion, degrading optical quality. Therefore, co-doping requires careful optimization to balance improved absorption with increased optical and thermal losses \cite{Liang2023SolarLasers}.

Ytterbium (Yb)-doped YAG provides markedly superior thermal performance compared to Nd:YAG, primarily due to its lower quantum defect. As a quasi-three-level gain medium, the broad emission bandwidth of Yb:YAG, enables pulse durations down to a few hundred femtoseconds. Near room temperature, its emission and absorption spectra partially overlap, and its reduced optical quenching allows for doping concentrations beyond \SI{10}{\%} \cite{Yang2002}. However, the energy gap between the lower laser level ($^4I_{13/2}$) and the ground state leads to significant thermal population, increasing the lasing threshold relative to Nd-based systems and requiring higher pump intensities. Moreover, the thermal population of the terminal laser level further reduces efficiency, intensifying thermal management challenges and causing thermo-optic distortions. To mitigate these effects and improve beam quality, \YbIII lasers have been realized in a variety of geometries, including fiber, thin-disk, and slab configurations \cite{snitzer1961proposed, martin1972multiple, eggleston1989deviation, giesen1994scalable, fattahi2014third, nubbemeyer20171, fattahi2016high, li20240, endo2007feasibility, wang2023investigation, petermann2005highly}.

In Yb:YAG thin-disk lasers, the active medium is water-cooled through one of its flat faces. The high surface-to-volume ratio enables efficient heat removal into the heat sink, resulting in a nearly uniform transverse temperature gradient within the homogeneously pumped central region of the disk. This minimizes stress-induced birefringence, thermal lensing, and aspherical aberrations, thereby preserving diffraction-limited beam quality. The cooled crystal face is often coated to function as a folding mirror or the end mirror of the resonator. To overcome the reduced pump absorption inherent to the thin gain medium, a multi-pass pumping scheme is employed in which multiple parabolic mirrors redirect pump light through the medium several times. The arrangement of parabolic mirrors determines the number of passes, which in turn directly affects both pump absorption efficiency and the resulting beam quality \cite{4244423}.

Moreover, \YbIII\ ions are well suited for radiation-balanced lasing \cite{nemova2021radiation}. In radiation-balanced lasers (RBL), the pump wavelength is red-shifted relative to the mean fluorescence wavelength of the gain medium, enabling anti-Stokes fluorescence cooling \cite{5440021}. In this process, energy in the form of phonons is transferred from the host material to the emitting ions, maintaining thermal equilibrium or cooling the gain medium. Under RBL conditions, the spontaneous fluorescence rate is comparable to the stimulated emission rate. This enables heat and entropy produced by optical pumping, arising from quantum defects and other nonradiative processes, to be dissipated via anti-Stokes fluorescence, thereby reducing excess heat below the quantum defect limit \cite{737628, 1035981, bowman1999lasers, mungan2003thermodynamics}.

Radiation cooling has been demonstrated in Yb:YAG slab lasers \cite{Lei2023, epstein2009, Nemova2021} and thin-disk lasers \cite{nemova2014thin, yang2019radiation, 8883199} by pumping the gain medium with a narrowband laser under RBL conditions \cite{epstein2009optical}. Recent studies further suggest that optical cooling can also be achieved using incoherent illumination, such as light-emitting diodes or spectrally filtered sunlight, with efficiencies comparable to those obtained with coherent, laser-based pumping \cite{ghonge2024photoluminescent}.

In this work, we investigate the feasibility of a solar-pumped Yb:YAG laser, with particular emphasis on determining whether radiation-balanced lasing can be achieved to support sustainable, self-cooling performance. We first analyze an advanced solar-pumped thin-disk architecture employing Yb$^{3+}$ as the gain medium. This design incorporates a water-cooled thin disk gain medium in combination with a solar-concentrating dome that traps and re-images the incident sunlight. The concentrator geometry is optimized to address challenges associated with the low spatial coherence of solar radiation, which typically limits the utility of conventional multi-pass pumping approaches.

Building on these insights, we then examine a solar-pumped, radiation-balanced Yb:YAG laser. In particular, we evaluate the impact of pump spectral bandwidth on both laser gain and radiative cooling, and we identify the boundary conditions necessary to sustain RBL operation. The proposed scheme enables efficient extraction of anti-Stokes fluorescence and supports RBL operation in both thin-disk and rod geometries. These studies outline promising directions for the development of thermally resilient, energy-efficient, and environmentally sustainable solar-pumped laser technologies.

\section*{Water-cooled solar-pumped thin-disk laser}
\begin{figure}[t]
    \centering
    \includegraphics[width=\linewidth]{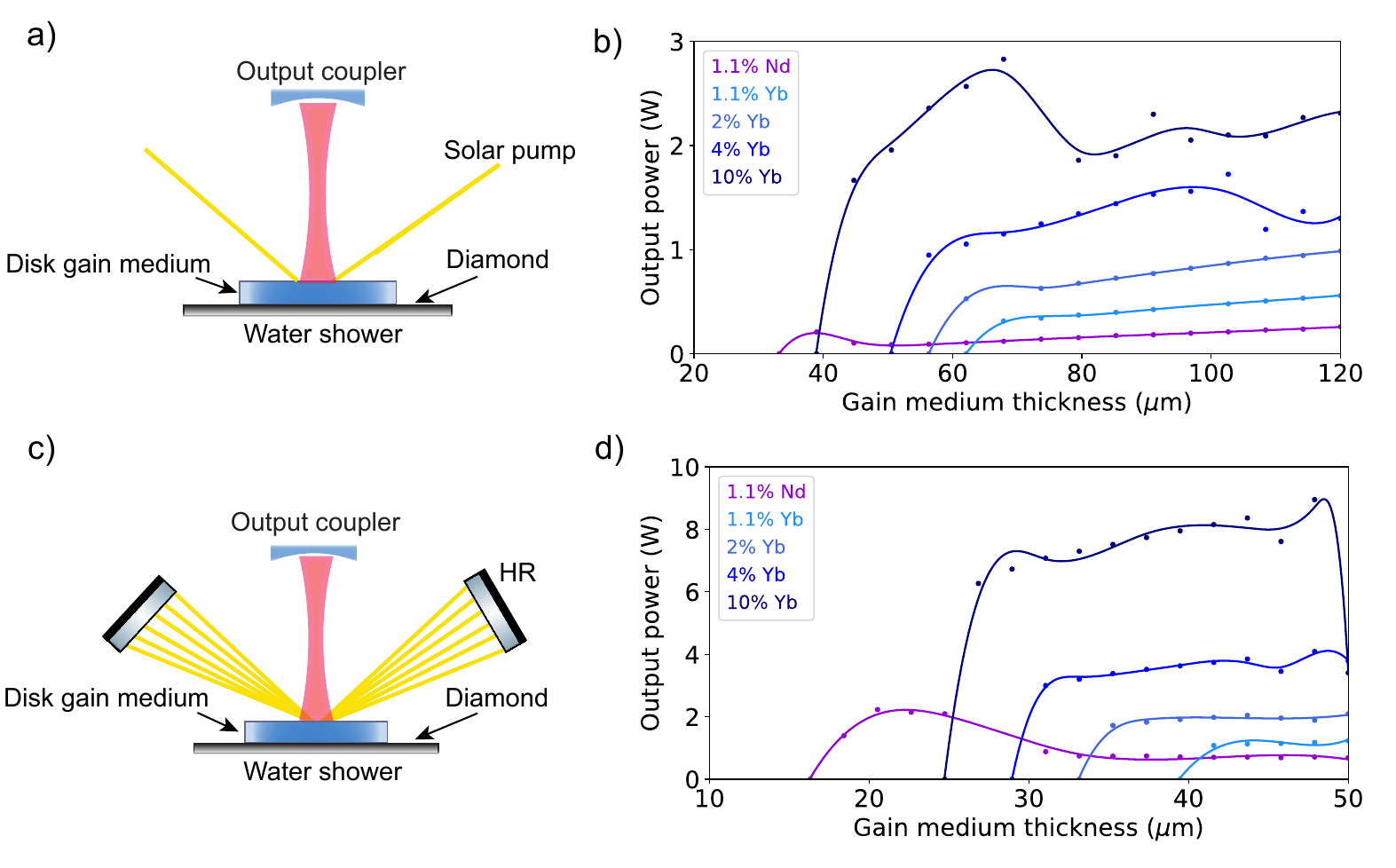}
    \caption{\textbf{Comparison of \NdIII{} and \YbIII{} gain media in a thin-disk geometry.} a) Schematic of a single-pass solar-pumped thin-disk laser. b) The output power versus gain medium thickness of the single-pass pumped laser cavity for a \SI{1.1}{\percent}-doped Nd:YAG and various doping concentrations from \SI{1.1}{\percent} to \SI{10}{\percent} for Yb:YAG thin-disk. c) Schematic of a six-pass solar-pumped thin-disk laser. d) The output power versus thickness of a six-pass solar-pumped laser with a \SI{1.1}{\percent}-doped Nd:YAG and various doping concentrations from \SI{1.1}{\percent} to \SI{10}{\percent} for Yb:YAG. The unfiltered solar irradiance of \SI{1367}{\watt\per\meter\squared} for Nd:YAG and the filtered solar irradiance of \SI{85.8}{\watt\per\meter\squared} for Yb:YAG are considered for both schemes.}
    \label{fig1}
\end{figure}

\begin{figure}[t]
    \centering
    \includegraphics[width=\linewidth]{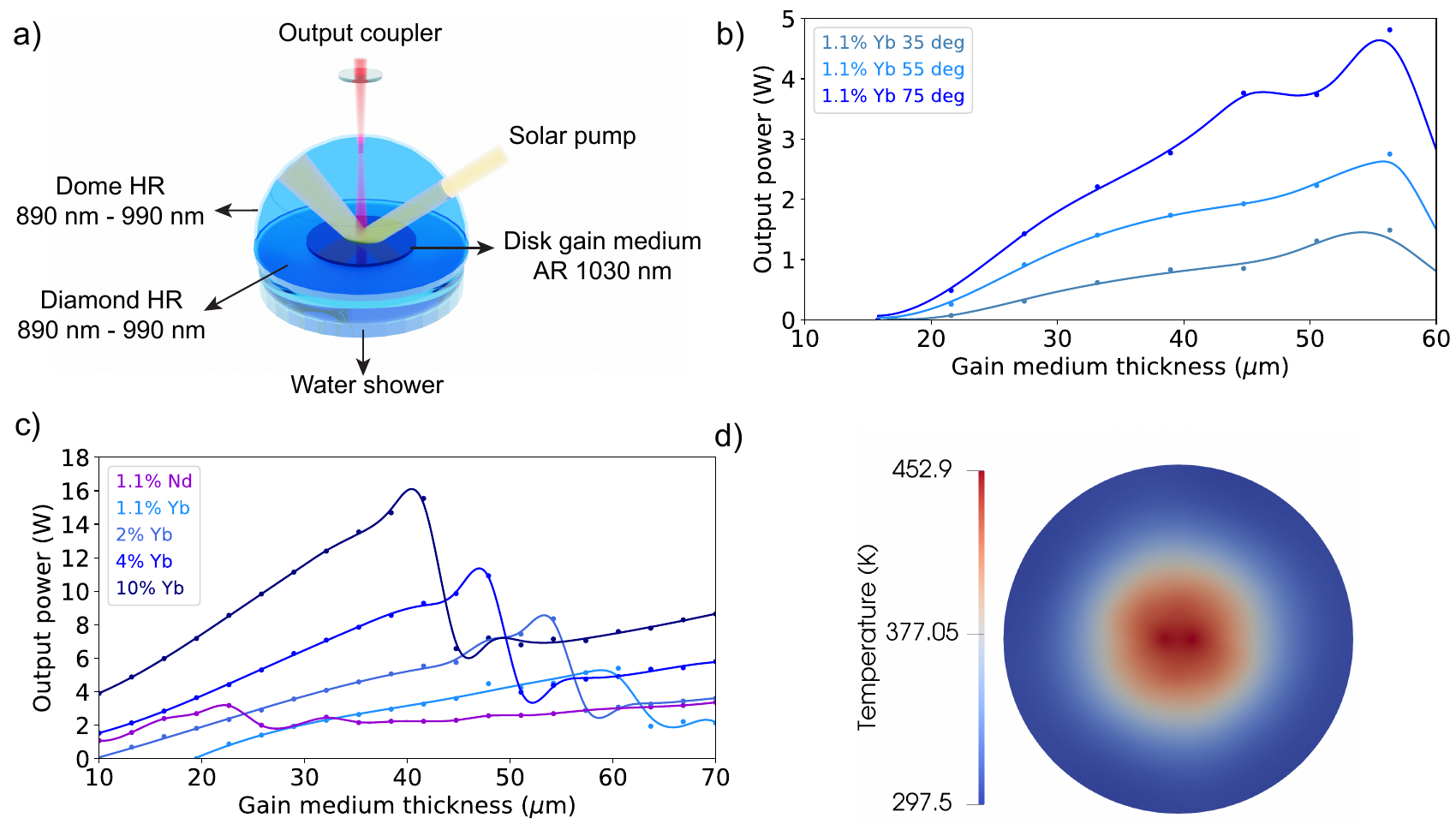}
    \caption{\textbf{Dome solar-pumped thin-disk laser.} a) Schematic of a dome solar-pumped thin-disk laser. b) The thickness of the crystal versus the output power of the laser for various solar incident angles. The calculation is performed for \SI{1.1}{\percent}-doped Yb:YAG gain medium. c) The output power of the dome concentrator concept versus the thickness of the gain medium. It is assumed that the solar beam enters the dome at an oblique angle of \SI{75}{\degree} relative to the disk surface. d) Temperature distribution on the thin-disk for a 10\% Yb-doped gain medium at a thickness of \SI{40}{\micro\meter} and solar pump incident angle of \SI{75}{\degree}. The accumulation of heat in the gain medium induces deformation of the thin disk. As shown in panel c, beyond this regime and for increased thickness, the laser output decreases significantly due to thermally induced curvature changes of the thin disk.}
    \label{fig2}
\end{figure}

Both \YbIII and \NdIII gain media were evaluated for implementation in a thin-disk geometry (see Methods). For scalability and consistent comparison, we assumed that \SI{1}{\square\meter} of sunlight is collected and concentrated onto the gain medium. Considering previously demonstrated solar-pumped Nd:YAG lasers, we assume that the unfiltered extraterrestrial solar spectrum, spanning \SIrange{200}{3000}{\nano\meter} and corresponding to a total irradiance of \SI{1367}{\watt\per\meter\squared}, is used to pump the Nd:YAG thin-disk gain medium. The resonator configuration for this system was designed to support lasing at \SI{1064}{\nano\meter}. For pumping of the Yb:YAG gain medium, we consider a spectrally filtered portion of the extraterrestrial solar spectrum spanning \SIrange{890}{990}{\nano\meter}, corresponding to a solar irradiance of \SI{85.8}{\watt\per\meter\squared}. The use of a spectrally filtered pump is intended to establish a direct comparison with the radiation-balanced lasing case discussed in the following section. The laser cavity for the Yb:YAG system is correspondingly optimized for emission at \SI{1030}{\nano\meter}.

The thin-disk gain medium was mounted on a diamond substrate that was cooled from below using a water shower, while the top and lateral surfaces of the disk remained uncooled and exposed to ambient air. The front surface of the gain medium was coated with a \SI{100}{\percent} anti-reflection layer, whereas the rear surface of the gain medium and the diamond surface were coated with a \SI{99.5}{\percent} reflective layer at both the pump and lasing wavelengths. Pumping was implemented in a quasi-end-pumped configuration, where the incident pump beam traverses the crystal at an oblique angle. To assess the degree of saturation in the gain medium and quantify the influence of the proposed integrated dome, we first examined the laser output power as a function of the number of controlled passes of solar light through the gain medium. Fig.~\ref{fig1}a depicts the cavity geometry corresponding to the single-pass solar-pumping configuration.

For Nd:YAG, a doping concentration of \SI{1.1}{\percent} was assumed, as this value is considered optimal due to the larger ionic radius of Nd\textsuperscript{3+} compared with Y\textsuperscript{3+}. Higher Nd\textsuperscript{3+} doping levels introduce significant lattice stress during crystal growth, which often leads to dislocations and other material defects~\cite{Kanchanavaleerat2004}. A fluorescence lifetime of \SI{235}{\micro\second} was used for the Nd:YAG simulations. Yb:YAG can accommodate Yb\textsuperscript{3+} doping levels up to approximately \SI{25}{\percent}, owing to the close similarity in ionic radii between Yb\textsuperscript{3+} and Y\textsuperscript{3+}~\cite{Crylaser, Qiu2002}. However, the doping concentration was varied only from \SI{1.1}{\percent} to \SI{10}{\percent}, as higher concentrations are known to induce concentration quenching~\cite{Lei2023, epstein2009, Yang2002, Nemova2021,yan2014gain}. A constant fluorescence lifetime of \SI{1}{\milli\second} was assumed over this doping range. Fig.~\ref{fig1}b presents the resulting output power for single-pass pumping at various gain-media thicknesses and doping concentrations. It is observed that Nd:YAG exhibits a lower lasing threshold than Yb:YAG at comparable doping levels, primarily due to its four-level energy level scheme. Yb:YAG requires roughly an order of magnitude higher doping to achieve a similar threshold. However, it provides up to three times greater output power.

A second series of simulations was performed to investigate the effect of re-imaging the residual reflected pump power for up to five additional passes, thereby enabling multi-pass pumping to compensate for the inherently low single-pass gain in the thin-disk geometry. Fig.~\ref{fig1}c illustrates the schematic of the six-pass pumping configuration, and Fig.~\ref{fig1}d presents the corresponding output-power scaling. The results show up to a twofold reduction in the lasing threshold across the range of doping concentrations and gain-medium thicknesses examined.

Solar light collected through a combination of primary and secondary focusing can be concentrated into a beam with a diameter below \(2\,\mathrm{mm}\), as demonstrated in numerous previous studies \cite{li2026generatingquantumentanglementsunlight, renene2023_03_006, brar2025solar, article2022_s2666523922001386, Kublbock.2023}. Efficient re-imaging of solar light using conventional mirror arrangements is challenging as the solar light is only partially coherent \cite{mashaal2012spatial, agarwal2015coherence}. To address this limitation, we developed a re-imaging concept tailored for solar-pumped thin-disk lasers (Fig.~\ref{fig2}a). The design incorporates a spherical half-dome with a diameter of \SI{25}{\mm} that confines and redirects solar radiation, enabling effective multi-pass pumping. For the simulations, a circular emitter with a diameter of \SI{2}{\mm} was placed on the dome surface. The inner surface of the dome was assumed to be coated with a \SI{100}{\percent} reflective layer at the pump wavelength, except in the emitter region. Rays impinging on the emitter were considered fully absorbed, reproducing the function of a \SI{2}{\mm} entrance aperture through which the solar beam enters the dome.

Since only a small fraction of the pump radiation is absorbed in each pass, multi-pass operation is achieved by allowing the pump beam to reflect off the rear surface toward the inner surface of the dome. This process continues iteratively until the beam is eventually redirected onto the emitter aperture and exits the dome (See Fig.1-SI in supplementary Information). This geometry, therefore, enables multiple effective pump passes while accommodating the limited spatial coherence of solar radiation. Fig.~\ref{fig2}b shows the dependence of the laser output on the incident angle of the solar pump beam on the gain medium. The angle is defined with respect to the surface of the thin disk. The results indicate that incident angles closer to normal incidence yield higher output power. This enhancement arises from improved spatial overlap between the pumped volume in the gain medium and the laser cavity mode. In addition, at larger incident angles, the effective irradiated area increases, thereby reducing the overlapped pump flux with the cavity mode.

As shown in Fig.~\ref{fig2}c, implementing the dome-based multi-pass pumping scheme lowers the lasing threshold for both gain media substantially compared to the other geometries. In addition, saturation is achieved at smaller gain-medium thicknesses for lower Yb:YAG doping concentrations, indicating more efficient pump utilization and improved cooling within the thin-disk configuration. The highest output power is obtained for a Yb:YAG gain medium with 10\% doping and a thickness of \SI{40}{\micro\meter}. Beyond this regime, the thermal load on the gain medium leads to surface deformation, thermal lensing, and mechanical stress within the crystal. Fig.~\ref{fig2}d illustrates the resulting thermal distribution and surface deformation along the xy-axes of the disk beyond \SI{40}{\micro\meter} thickness.

\section*{Radiation-balanced solar-pumped thin-disk laser}

\begin{figure}[t!]
    \centering
    \includegraphics[width=\textwidth]{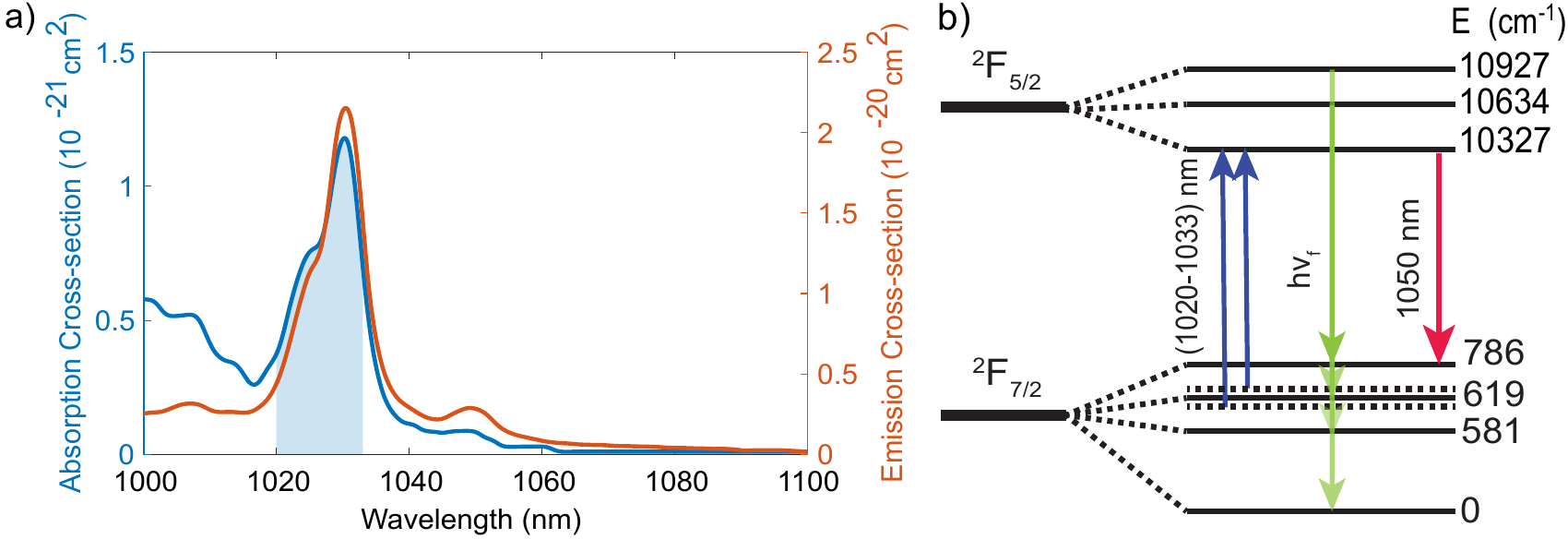}
    \caption{\textbf {Required properties of \YbIII ions for radiation balanced lasing.} a) Absorption and emission cross section of $\mathrm{Yb}^{3+}$. The blue shaded area shows the filtered spectral bandwidth in the solar spectrum from \SI{1020}{\nano\meter} to \SI{1033}{\nano\meter} to pump the Yb:YAG gain medium for radiation-balanced operation \cite{5440021}. b) Energy level diagram of the Yb:YAG gain medium in the radiation balance regime. The blue arrows show the pump spectrum, the red arrow is the lasing wavelength, and the green arrows ($\nu_f$) refer to the anti-Stokes emission \cite{8883199, phdthesis}. }
    \label{fig3}
\end{figure}

\begin{figure}[ht]
    \centering
    \includegraphics[width=1\linewidth]{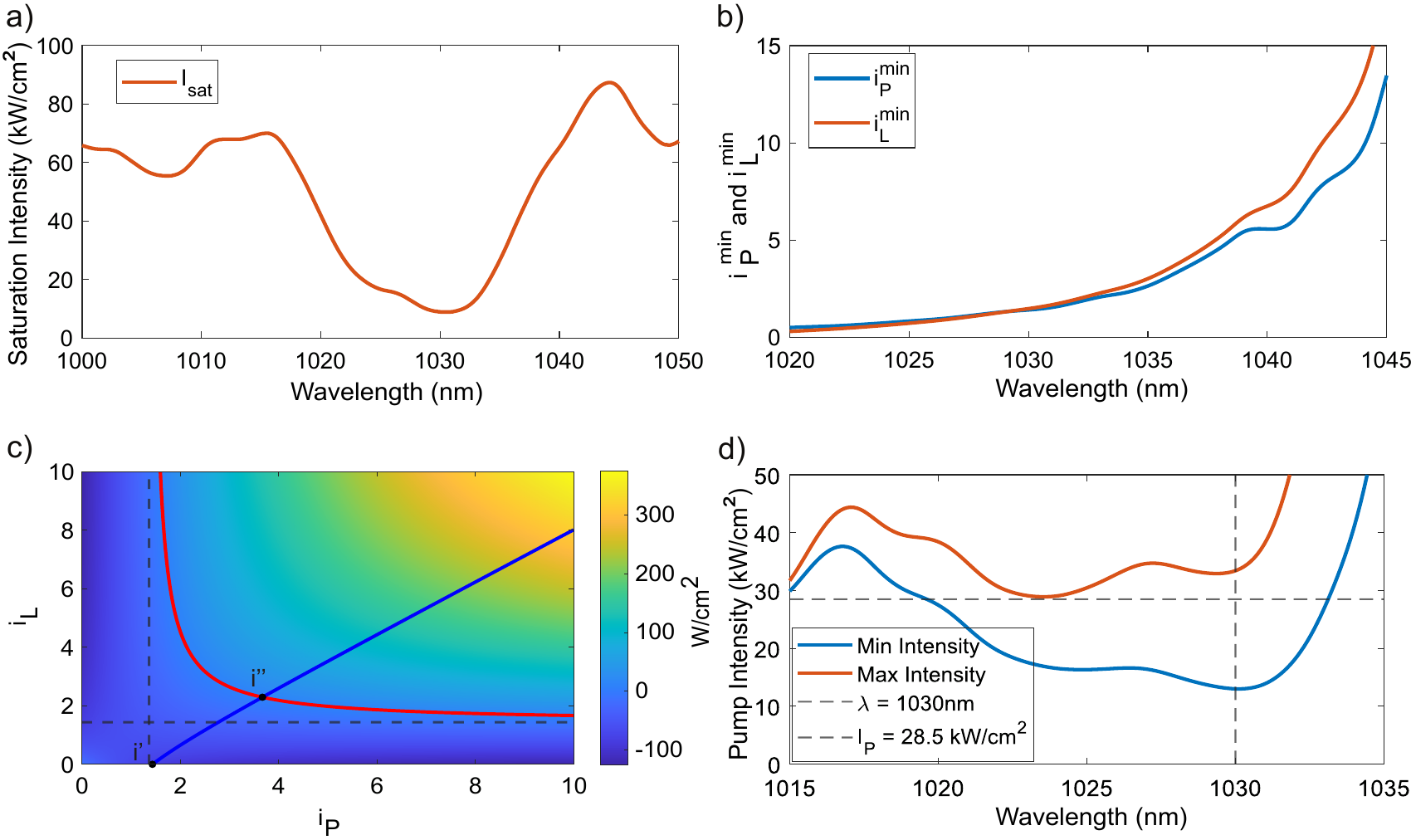}
    \caption{\textbf {Required solar spectrum and solar intensity for radiation balanced lasing.} a) Wavelength-dependent required intensity to saturate the atomic transition in Yb:YAG gain medium. b) The minimum pump ($i^{\min}_P$) and lasing ($i^{\min}_L$)intensities required to achieve RBL for various pump wavelengths. c) Thermal power density map for pump wavelength  $\lambda_P = \SI{1030}{\nm}$ and lasing wavelength $\lambda_L = \SI{1050}{\nm}$. The Red curve represents the exact condition for RBL, and the blue curve illustrates the relationship between normalized pumping and lasing intensities. For this simulation, the crystal length of \SI{1}{\milli\meter} with a doping concentration of 5\%\cite{5440021} and cavity loss of 0.5\% is assumed. d) RBL range is defined for solar spectral bandwidth of \SI{1020}{\nano \meter} to \SI{1033}{\nano \meter}. The blue curve indicates the threshold pump intensity. The orange curve shows the maximum allowed pump intensity for operation in RBL.}
    \label{fig4}
\end{figure}

During laser operation, optical pumping at photon energy $h\nu_{P}$ establishes a population inversion, which subsequently leads to stimulated emission at the lower photon energy $h\nu_{L}$. The net thermal load in the gain medium arises from the difference between the absorbed pump power and the total emitted optical power. The latter includes both stimulated emission at $\nu_{L}$ and spontaneous emission (fluorescence) at $\nu_{f}$. In conventional lasers, anti-Stokes fluorescence is negligible because of its comparatively small emission cross section, resulting in a predominantly exothermic process where heat accumulates in the gain medium. However, when operating under the anti-Stokes--dominated condition $h\nu_{L} < h\nu_{P} < h\nu_{f}$, radiative processes can offset or even surpass the heat generated by nonradiative mechanisms, enabling athermal or radiation-balanced (self-cooling) laser operation (see Methods).

As previously noted, Yb:YAG possesses the requisite properties for operation as a cooling-grade gain medium, with a mean fluorescence wavelength of \SI{1010}{\nano\meter} \cite{brown2011ybyag, wu2023influence}. Fig.~\ref{fig3}a presents the absorption and emission cross sections of Yb:YAG within the spectral region relevant to radiation-balanced lasing near \SI{1050}{\nano\meter}. To satisfy the anti-Stokes pumping condition, the pump wavelength must exceed the mean fluorescence wavelength, i.e., $\lambda_{P} > \lambda_{f} = \SI{1010}{\nano\meter}$ (see Fig.~\ref{fig3}b). Both the absorption and emission cross sections exhibit significant wavelength-dependent variation across this region. Consequently, the required fractional excitation, as well as the minimum pump and laser intensities necessary for athermal operation, are strongly influenced by the choice of pump spectral bandwidth.

To determine an appropriate solar pumping spectrum capable of supporting radiation-balanced lasing in Yb:YAG, numerical simulations were performed. Fig.~\ref{fig4}a illustrates the dependence of the saturation pump intensity on the spectral bandwidth of the solar irradiation considered suitable for radiation-balanced operation. At \SI{1030}{\nano\meter}, where both the absorption and emission cross sections reach their maximum values, achieving transition saturation requires a pump saturation intensity of approximately \SI{9}{\kilo\watt\per\centi\meter\squared} within the gain medium (see eq. \ref{eq6} in Methods). %To identify the most efficient spectral bandwidth for pumping at RBL, the power spectral density for one sun based on black body radiation at the temperature of $5778\,\mathrm{K}$ is considered. To reach the saturation intensity of \SI{9}{\kilo\watt\per\centi\meter^2} at \SI{1030}{\nano\meter}, the intensity of  $85.715\times10^6$ Suns is required.

Fig.~\ref{fig4}b displays the wavelength dependence of the minimum pump and intra-cavity lasing intensities (see eq. \ref{eq4} in Methods). Both values remain relatively close within the spectral range of \SI{1020}{\nano\meter} to \SI{1033}{\nano\meter}, but begin to diverge at longer wavelengths, indicating the need for higher intra-cavity laser intensity to meet the RBL condition. Figure~\ref{fig4}c illustrates the relationship between pump intensity and intra-cavity laser intensity for a pump wavelength of \SI{1030}{\nano\meter}. The red curve represents the condition of zero net heat deposition in the gain medium, defining the boundary for radiation-balanced operation. The blue curve shows the pump--laser intensity relationship under steady-state lasing for single-pass pumping.

Two key operating points appear as black markers on the blue curve. The lower marker (\(i'\)) indicates the lasing threshold at $i_{P} = 1.44$, corresponding to the minimum pump intensity required to sustain stimulated emission. The upper marker (\(i''\)) identifies the point at which the blue curve intersects the red curve, occurring at $i_{P} = 3.67$ and an intra-cavity intensity of $i_{L} = 2.29$. At this intersection, all heat generated in the crystal is fully compensated by anti-Stokes fluorescence, marking the onset of radiation-balanced lasing and establishing the maximum pump intensity that maintains cooling performance.

For stable radiation-balanced operation, the system must operate at pump and intracavity lasing intensities that lie on the red curve. When the pump intensity exceeds the lasing threshold but remains below the radiation-balanced limit, the gain medium undergoes net cooling. Above the radiation-balanced boundary, lasing may still persist; however, the net heat deposition becomes positive, leading to heating of the gain medium. The accompanying colour map further illustrates the net heat deposition as a function of pump and intra-cavity laser intensities, distinguishing cooling-enabled conditions (below the red curve) from net-heating regions (above the red curve).

\begin{figure}[t]
    \centering
    \includegraphics[width=1\linewidth]{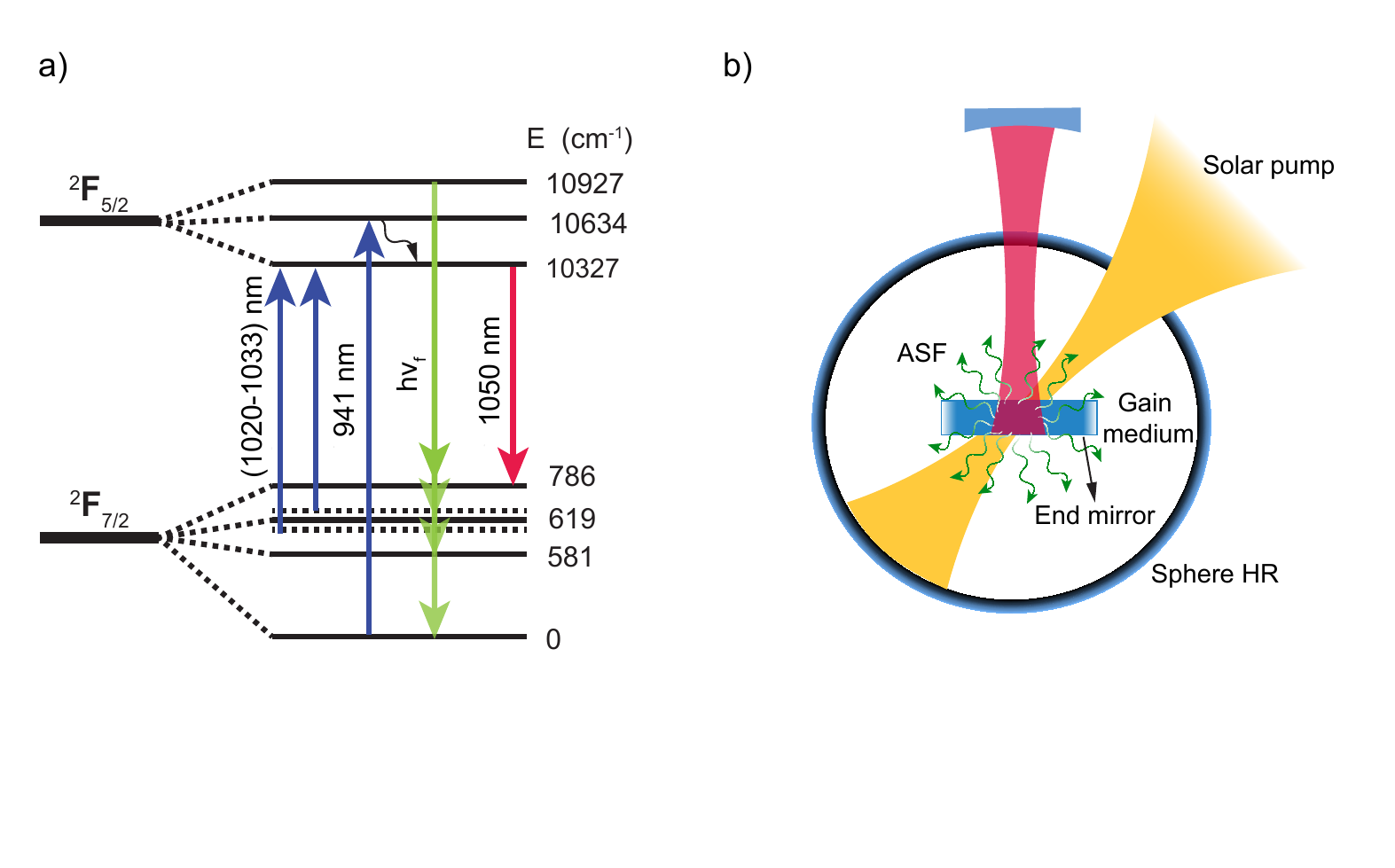}
    \caption{\textbf {Concept of a solar-pumped Yb:YAG laser including a sphere concentrator operating in the radiation-balanced regime.} a) Radiation-balanced solar-pumped laser operation can be achieved using a dual-pump configuration. While the gain medium is pumped at the radiation-balanced regime, an additional pump at \SI{941}{\nano\meter}, corresponding to a non-zero phonon transition, is introduced. The resulting balance between heat generation and optical refrigeration enables radiation-balanced operation under solar pumping. b) A spherical concentrator collects and refocuses sunlight onto the gain medium to increase pump absorption, while also enabling efficient extraction of anti-Stokes fluorescence (ASF) in all directions to support radiative cooling. The inner surface of the concentrator is coated for high reflectivity within the \SI{1020}{\nano\meter}--\SI{1033}{\nano\meter} spectral band required for RBL operation, excluding the area for solar input.}
    \label{fig5}
\end{figure}

These operational limits were evaluated for pump wavelengths longer than the mean fluorescence wavelength, while maintaining a fixed lasing wavelength of \SI{1050}{\nano\meter}. Figure~\ref{fig4}d summarizes the minimum and maximum pump intensities required for radiation-balanced operation when solar pumping is confined to the \SI{1015}{\nano\meter}--\SI{1035}{\nano\meter} spectral range. The most favorable performance occurs between \SI{1020}{\nano\meter} and \SI{1033}{\nano\meter}, where the RBL condition is satisfied at a pump intensity of \SI{28.5}{\kilo\watt\per\centi\meter\squared}.

Within this wavelength range, the available solar power is approximately \SI{9.5}{\watt\per\meter\squared}. When concentrated onto a \SI{2}{\milli\meter} diameter spot, this yields a peak intensity of only \SI{300}{\watt\per\centi\meter\squared}. Achieving the required \SI{28.5}{\kilo\watt\per\centi\meter\squared} would therefore require increasing the solar collection area by orders of magnitude, which is neither practical nor economically viable. As shown in Figure~\ref{fig4}c, the lasing threshold is not reached at this pump intensity in a single pass pumping geometry. However, in a multipass pumping configuration with six passes through a \SI{1}{\milli\meter}-thick Yb:YAG gain medium, the lasing threshold can be achieved (see Figure 2-SI in supplementary information). Figure~\ref{fig4}c indicates that at this pump intensity, the laser operates in the net cooling regime. To enable stable operation at a desired temperature or radiation-balanced point, an additional heating mechanism can be introduced by taking advantage of the broadband nature of the solar radiation. For instance, solar light at \SI{941}{\nano\meter} with \SI{2}{\nano\meter} spectral bandwidth can be spectrally filtered and used to simultaneously pump the gain medium. The heat generated by a single-pass pump at this wavelength compensates for the cooling induced by the primary pump and stabilizes the temperature of the gain medium. Therefore, the RBL condition can be achieved using dual-wavelength pumping: six-pass pumping over a spectral bandwidth of \qtyrange{1020}{1033}{\nano\meter}, combined with single-pass pumping at \qty{941}{\nano\meter}.

Fig.~\ref{fig5} illustrates a proposed configuration designed to enable RBL operation under multi-pass solar pumping. In the absence of external thermal management requirements, a fully spherical concentrator geometry can be employed to efficiently re-image and confine sunlight onto the gain medium. Radiative heat extraction further permits an increase in the gain-medium thickness, enabling gain saturation without compromising thermal performance. The spherical geometry facilitates isotropic escape of anti-Stokes fluorescence, thereby minimizing photon reabsorption and reducing the likelihood of radiative trapping. The gain medium can be implemented in either rod or thin-disk geometries, depending on design constraints. A solar entrance filter selects the pump light at two desired wavelengths with appropriate spectral bandwidths. The inner surface of the concentrator is coated for high reflectivity across the solar pump band spanning \SI{1020}{\nano\meter} to \SI{1033}{\nano\meter}. One surface of the gain medium is coated to act as the cavity end mirror with high reflectivity at \SI{1050}{\nano\meter}, while the remaining surfaces are anti-reflection coated at the solar pump wavelengths to maximize coupling efficiency. It is important to note that, although lasing at \SI{1050}{\nano\meter} is demonstrated here, the inverse configuration is also feasible. For example, a cavity operating at \SI{1030}{\nano\meter} can be pumped at \SI{941}{\nano\meter} while utilizing the \SIrange{1020}{1033}{\nano\meter} spectral band for radiative cooling. Alternatively, a dual-wavelength cavity supporting simultaneous lasing at both \SI{1030}{\nano\meter} and \SI{1050}{\nano\meter} could be implemented to leverage both pump channels. Finally, we emphasize that the present estimates do not account for population dynamics under dual-wavelength pumping. Since the \SI{941}{\nano\meter} pump contributes to populating the excited state, it is expected to reduce the lasing threshold, which would be highly advantageous for the proposed scheme.
%Furthermore, the intensity for each wavelength, with a FWHM of \SI{0.5}{\nano\meter}, is \SI{33.4}{\watt\per\centi\meter^2}. To reach the \SI{28.5}{\kilo\watt\per\centi\meter^2} required pump intensity for the RBL operation, the collection area should be increased or, as demonstrated in this work, multi-pass pumping can be employed \cite{Kabelac2012, Markvart2008, 2009Pierre, DelgadoBonal2017}.
\section*{Conclusion}

Solar-pumped lasers offer a compelling pathway toward renewable laser energy conversion and space-based photonic applications, yet their performance has long been constrained by thermal loading and limited power scalability. Here, we propose the concept of a solar-pumped thin-disk laser employing both Nd:YAG and Yb:YAG gain media. The thin-disk geometry enables efficient heat extraction, significantly suppressing thermal gradients and surface deformation; however, it suffers from low solar pump absorption per pass. We numerically demonstrate that while Nd:YAG benefits from a lower lasing threshold under single- and six-pass pumping compared to Yb:YAG due to its four-level energy scheme, this advantage becomes marginal when combined with the dome concentrator multi-pass pumping concept introduced in this work. The dome concentrator enhances solar photon trapping and absorption efficiency, enabling both gain media to reach comparably low lasing thresholds. Under these conditions, Yb:YAG clearly outperforms Nd:YAG, offering superior thermal management and markedly improved power scalability. The Nd:YAG doping limit of approximately $1.1\%$, imposed by fluorescence quenching, severely constrains output power. In contrast, thin-disk Yb:YAG supports substantially higher doping concentrations, up to $10\%$, enabling output powers up to three times higher than those achievable with Nd:YAG.

We identify Yb:YAG as a promising candidate for a solar-pumped radiation-balanced lasing, in which anti-Stokes fluorescence, combined with dual-wavelength pumping, compensates for temperature variations during lasing and enables sustainable operation. Our calculations indicate that RBL conditions can be achieved at a solar pump intensity of approximately \SI{28.5}{\kilo\watt\per\centi\meter\squared} within a narrow spectral window from \SI{1020}{\nano\meter} to \SI{1033}{\nano\meter}. This requirement implies that fully radiation-balanced operation under direct solar pumping necessitates a large collection area. However, the spherical multi-pass concentrator proposed here provides an effective route to reach the lasing threshold even at significantly lower pump intensities (~\SI{300}{\watt\per\centi\meter\squared}) by enhancing solar photon trapping while maintaining a compact system footprint. In this regime, the system operates in the net cooling domain. Radiation-balanced operation can then be restored by introducing controlled heating of the gain medium via off-resonant pumping at \SI{941}{\nano\meter}, corresponding to a non-zero phonon transition. This approach enables the use of thicker gain media without compromising thermal performance. At the same time, the spherical geometry promotes isotropic escape of anti-Stokes fluorescence, thereby reducing photon reabsorption and radiative trapping. Both rod and thin-disk laser architectures are compatible with this concept.

These results establish Yb:YAG lasers combined with sphere-concentrator multi-pass solar pumping as a promising platform for compact, scalable, and high-efficiency solar-pumped laser systems. Future investigations may leverage Purcell-enhanced cavity designs to reduce radiation-balanced laser threshold requirements further. This work lays the groundwork for sustainable photonic technologies with applications in space-based power generation, remote sensing, and advanced laser energy systems. 

\section*{Methods}

\begin{figure}[htbp]
    \centering
    \includegraphics[width=1\linewidth]{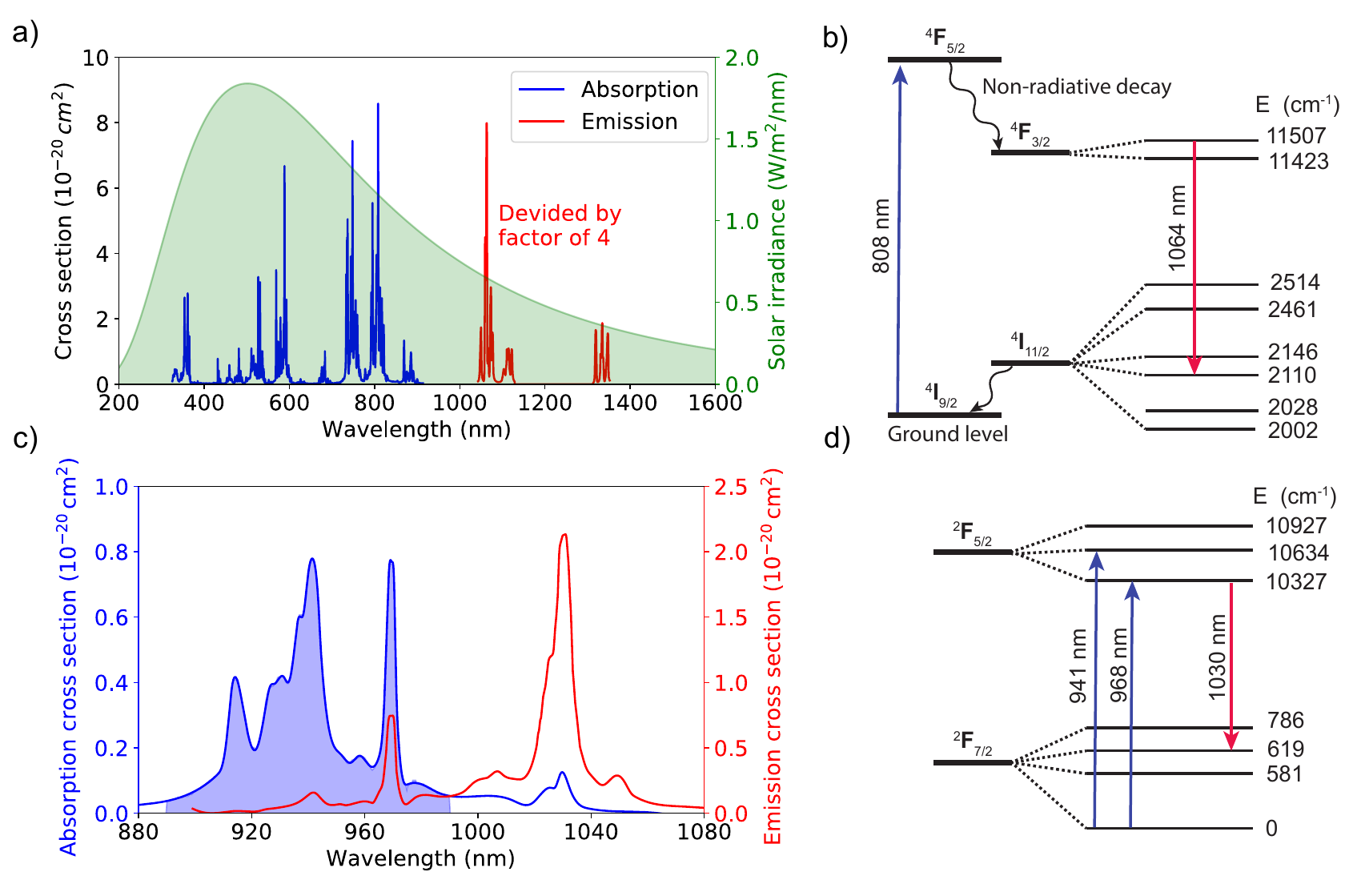}
    \caption{\textbf {Lasing properties of \NdIII and \YbIII ions.} a) Absorption and emission cross sections of \NdIII overlayed with extraterrestrial solar irradiance \cite{Yao.2015,10.1117/12.737627}. b) Energy level diagram of the Nd:YAG gain medium \cite{Liang.2023,Deng.2018,Xu.2019}. c) Absorption and emission cross-section of \YbIII. The blue shaded area shows the spectral bandwidth of the absorption cross section of \YbIII, which is filtered in the solar spectrum to pump the \YbIII gain medium \cite{5440021}. d) Energy level diagram of the Yb:YAG gain medium \cite{phdthesis}.}
    \label{fig6}
\end{figure}

Numerical simulations were performed using the ``Advanced software for laser design'', which considers the superposition and competition of laser cavity modes via dynamic mode analysis \cite{Wohlmuth09, wohlmuth2010dynamic}. In dynamic mode analysis, rate equations are solved for each excited transverse mode in the cavity. The optical wave is then described as a dynamic superposition of several eigenmodes, allowing for the precise calculation of the laser output power and beam quality. The deformation of the gain medium, stress, and thermal effects are calculated via finite element analysis, and their impact on thermal lensing and the stability of the laser resonator is considered. Fig.~\ref{fig6} shows the absorption and emission cross sections and energy level diagram of \NdIII\ and \YbIII\ ions. The extraterrestrial solar spectrum from \SI{200}{nm} to \SI{3000}{nm} corresponding to \SI{1367}{\W \per \meter\squared} solar irradiance was used to pump Nd:YAG gain medium, due to its broad absorption bandwidth. The laser cavity for Nd:YAG was designed for lasing at \SI{1064}{nm}. 

For pumping Yb:YAG gain medium, the filtered extraterrestrial solar spectrum with the bandwidth of \SI{890}{nm} to \SI{990}{nm} is used. The solar irradiance in this spectral bandwidth corresponds to \SI{85.8}{\watt\per\meter\squared}. The laser cavity for Yb:YAG is designed for lasing at \SI{1030}{nm}. It is assumed that the solar beam is concentrated to a \SI{2}{\milli\meter} diameter spot using a primary and secondary optical collection system \cite{fan2020design, wang2019review}. This region is considered fully transparent to solar radiation, allowing incident sunlight to enter the dome and reflected light to escape. \num{500000} rays  were used to simulate the solar light via ray tracing. A disk-shaped gain medium with a diameter of \SI{4}{mm} mounted on a \SI{0,5}{mm}-thick diamond substrate with a \SI{25}{mm} diameter is assumed for both gain media \cite{speiser2016}. The diamond is cooled by a water shower from beneath. The top and the side surfaces of the gain medium are not cooled and are in ambient air under atmospheric pressure. The water cooling of the gain medium is calculated via 
\begin{equation}
    k \cdot \diff{T}{n}- \beta \cdot (T - T_{\text{cool}})\big|_\Gamma = 0
\end{equation}
where $\beta$ is the heat transfer coefficient. When cooling with a water shower, \SI{0.05e-2}{\W \per \mm \per\K} for $\beta$ and  \SI{6}{\m \per \s} for flow rate is assumed. We assumed a cooling temperature of \SI{20}{\celsius} and a thermal conductivity of \SI{11}{\watt\per\metre\per\kelvin} for the YAG crystal.

It is assumed that the front surface of the gain medium and the diamond substrate have a \SI{100}{\%} anti-reflection coating, while the back surface of the gain medium has a \SI{99,5}{\%} reflective layer at the pump and lasing wavelengths. The gain medium is pumped in a quasi-end-pumped scheme, where the pump beam passes through the crystal under an oblique angle. The designed laser cavity for all the simulations in this section has a \SI{25}{mm} length comprising a \SI{99}{\%} output coupler in a Fabry P\'{e}rot cavity. This cavity design corresponds to the mode diameter of \SI{710}{\micro \m} on the flat thin-disk gain medium (see Fig.~\ref{fig1}a).  

For RBL operation, the following formulation is considered. As derived in \cite{5440021,8883199, bowman1999lasers}, during the lasing process, a pump laser with photon energy $h\nu_P$ creates population inversion, followed by stimulated emission at $h\nu_L < h\nu_P$. The net heat power deposited in the gain medium is the difference between the absorbed pump power and the emitted optical power. The latter is a combination of stimulated emission (laser) and spontaneous emission (fluorescence) at frequency $\nu_f$:

\begin{equation}
  H= \dot n_P h\nu_P - \dot n_L h\nu_L - \dot n_f h\nu_f.
\end{equation}

Here, $\dot n_P$ is the rate of photon absorption by the pump, $\dot n_L$ is the stimulated emission rate, and $\dot n_f$ is the spontaneous emission. Anti-Stokes fluorescence can be ignored for laser operation since $h\nu_P > h\nu_f$, as is the case in nearly all lasers. In this context, $\text{H} > 0$ represents an exothermic process, where heat is generated in the laser medium. In the anti-Stokes dominant condition given by $h\nu_L < h\nu_P < h\nu_f$, one obtains $\text{H} \leq0$  corresponding to an “athermal” (H = 0) or a self-cooling ($\text{H} < 0$) laser. The condition for a laser to operate at a radiation balance regime is:

\begin{equation}
    \frac{i_{P}^{\text{min}}}{i_P}+ \frac{i_{L}^{\text{min}}}{i_L} =1.
    \label{eq3}
\end{equation}

Here the minimum pump $i_{P}^{\text{min}}$ and minimum intra-cavity laser intensities $i_{L}^{\text{min}}$ for achieving the radiation balance are defined by

\begin{equation}
    i_{P}^{\text{min}} = \frac{\beta_{L}}{\beta_{P}-\beta_{L}}\frac{\nu_C-\nu_L}{\nu_P-\nu_L},
\quad
     i_{L}^{\text{min}} = \frac{\beta_{P}}{\beta_{P}-\beta_{L}}\frac{\nu_C-\nu_P}{\nu_P-\nu_L},
    \label{eq4}
\end{equation}

where the fractional excitation required to bleach the transition at $\lambda_i$ relative to its absorption $\sigma_{\text{abs}}$ and emission $\sigma_{\text{ems}}$ cross section is defined as:
\begin{equation}
    \beta_i = \frac{\sigma_{\text{abs}}(\lambda_i)}{\sigma_{\text{abs}}(\lambda_i)+\sigma_{\text{ems}}(\lambda_i)}, 
\quad
    \beta_{PL} = \frac{\beta_{P}\beta_L}{\beta_P-\beta_L}, 
\end{equation}

and $\nu_C$ is defined as the zero crossing frequency.

Saturation intensities are defined as 
\begin{equation}
    I_{Psat} \equiv \frac{h\nu_P}{\tau_F ({\sigma_{\text{abs}}(\lambda_P)+\sigma_{\text{ems}}(\lambda_P)})}, 
\quad
    I_{Lsat} \equiv \frac{h\nu_L}{\tau_F ({\sigma_{\text{abs}}(\lambda_L)+\sigma_{\text{ems}}(\lambda_L)})},
    \label{eq6}
\end{equation}

%\begin{equation}
 %   I_{\text{Psat}}= \frac{hc\beta_P}{\lambda_P \sigma_P \tau_F},
%\quad
 %   I_{\text{Lsat}}= \frac{hc\beta_L}{\lambda_L \sigma_L \tau_F},
  %  \label{eq6}
%\end{equation}

where $\tau_F$ is the fluorescence lifetime. 

The optimum value for intra-cavity lasing intensity is defined as \cite{8883199}: 
\begin{equation}
    i_L=\left(i_P+1\right)\left(\sqrt{\frac{1}{\xi}\cdot\frac{i_P \left(\frac{\beta_P}{\beta_L}-1\right)-1}{i_P+1}}-1\right).
\end{equation}

\section*{DECLARATIONS}
This work was supported by research funding from the Max Planck Society. The authors do not declare any competing interests. The data supporting this study's findings are available from the corresponding author upon reasonable request.
%\section{Supplementary Information}
%A parameter study of the crystal length was conducted in ASLD, where a \SI{25}{\milli\meter} cavity was designed with an output coupler of radius \SI{1}{\meter} and a reflectivity of 99.7\%. The radius of the crystal was set to \SI{2}{\milli\meter} and the radius of the pump on the crystal was \SI{1}{\milli\meter}. The bandwidth of the pump was taken from \SI{1020}{\nano\meter} to \SI{1033}{\nano\meter} with a input power of \SI{9.5}{\watt}. Different doping concentrations were also examined as shown in Fig:\ref{fig:ASLD_cooling}.
%\begin{figure}[htbp]
    %\centering
    %\includegraphics[width=0.75\linewidth]%{Pictures/SolarRBL/YbYAG_R2mm_1020nm_1033nm_9W5_smooth.pdf}
  %  \caption{Output power with varying length of the crystal.}
 %   \label{fig:ASLD_cooling}
%\end{figure}
%The output power has a linear dependence on the length of the crystal. Increasing the doping concentration also increases the maximum output power of the laser.
%\subsection{output coupler optimization}

%\begin{figure}
    %\centering
    %\includegraphics[width=0.9\linewidth]{Pictures/YbYAG/OC dome %Reflections.pdf}
    %\caption{Output coupler study for 1.1\% Yb and Nd for the angles of %75 degrees and 45 degrees. }
 %   \label{fig:enter-label}
%\end{figure}

 % \subsection{deformation and thermal}
% \begin{figure}[!ht]
%     \centering
%     \includegraphics[width=0.7\textwidth]{Pictures/SolarRBL/Crystal_length_crash_point_study.jpg}
%     % \caption{Yb:YAG crash point study from crystal length with Pump power of 2300W solar light}
%     % \label{fig:Crystal_length_shrash_point}
% \end{figure}
\bibliography{main}
\end{document}